# Non-line-of-sight imaging off a Phong surface through deep learning

Chen Zhou[1], Cheng-Yu Wang[1], Zhiwen Liu[*]

[1]These authors contributed equally
Department of Electrical Engineering, The Pennsylvania State University, University Park, PA, 16802, USA. *email: zzl1@psu.edu, N-211 Millennium Science Complex, The Pennsylvania State University, University Park, PA 16802

**A deep learning based non-line-of-sight (NLOS) imaging system is developed to image an occluded object off a scattering surface. The neural net is trained using only handwritten digits, and yet exhibits capability to reconstruct patterns distinct from the training set, including physical objects. It can also reconstruct a cartoon video from its scattering patterns in real time, demonstrating the robustness and generalization capability of the deep learning based approach. We further show that the presence of specular contribution plays a critical role in improving the image reconstruction quality. Several scattering surfaces with varying degree of Lambertian and specular contributions were examined experimentally; it is found that for a Lambertian surface the structural similarity index (SSIM) of reconstructed images is about 0.63, while the SSIM obtained from a scattering surface possessing a specular component can be as high as 0.93. To study the specular contribution enhanced reconstruction, a forward model of light transport was developed based on the Phong scattering model. Scattering patterns from Phong surfaces with different degrees of specular contribution were numerically simulated. These simulated patterns were then used to train and test deep neural nets. It is found that a specular contribution of as small as 5% can enhance the SSIM from 0.83 to 0.93, consistent with the results drawn from experimental data. To gain further insight, singular value spectra of the underlying transfer matrix were calculated for various Phong surfaces. As the weight and the shininess factor increase, i.e., the specular contribution increases, the singular value spectrum broadens and the 50-dB bandwidth is increased by more than 4X with a 10% specular contribution, which indicates that at the presence of even a small amount of specular contribution the NLOS measurement can retain significantly more singular value components, leading to higher reconstruction fidelity. With an ordinary camera and incoherent light source, this work enables a low-cost, real-time NLOS imaging system without the need of an explicit physical model of the underlying light transport process.**

Non-line-of-sight (NLOS) optical imaging aims to image hidden objects occluded from the direct view. Retrieval of image signals from such systems offers the prospect of monitoring hazardous environments, improving computer vision for navigation system, and enhancing search and rescue. The difficulties of NLOS imaging arise from the fact that the imaging system cannot directly detect the light emanating from the objects of interest, which are outside of the line of view, and instead can only rely on indirect measurement of scattered light from the environment, yielding distorted or partial information of the objects. As such, there exist significant challenges in modeling the light transport process, resolving information-carrying photons from noise, and reconstructing the hidden objects from measured data.

Transient imaging based gating methods or time-of-flight techniques have been introduced to realize NLOS imaging, which correlate object locations and topography with the arrival times of photons bounced off the object.[1–10] In order to achieve high temporal resolution and hence high-fidelity reconstruction, ultrafast pulsed lasers and high-speed, high-sensitivity detectors (e.g., streak camera and single-photon avalanche detector) are usually used, leading to bulky and costly system design. The corresponding physical models often involve estimate of the properties of hidden scenes using computationally expensive resources.

Passive imaging without laser illumination and time-resolved detection has been proposed by utilizing the computational periscopy technique.[11,12] This method uses an occluder with known shape and size to provide shadows and penumbrae and a camera positioned in best-conditioned regions to take multiple shots, and can achieve centimeter level accuracy. It relies on estimated light-transport model with heavy computation. Learning-based image reconstruction through random scattering media or multimode fibers has also been demonstrated, which utilizes machine learning trained with large amount of data to learn the light transport process.[13–19] Coherent light sources are usually used, with access to both intensity and phase (e.g., in the form of speckle) measurements of the scattered light. With sufficient training, the learning algorithm can generalize and approximate the light-transport process.[20,21]

Here we demonstrate a deep learning based NLOS imaging system off a scattering surface under incoherent light illumination. Unlike its coherent counterpart the phase information of the scattered optical field is not accessible in this case. We examine several scattering surfaces with varying Lambertian and specular contributions, and show that the presence of a specular component, albeit small, plays a surprisingly critical role in such a system and can dramatically enhance the reconstruction quality. We hypothesize that this small specular component can retain sufficient information of the primary light field associated with the hidden object to result in high-fidelity reconstruction, which is supported by our experimental studies and theoretical model. Trained only with handwritten digits from MNIST[22], our system can also reconstruct new patterns, such as letters, cartoon images, and real objects-wooden number blocks. This method incurs much less instrumental and computational costs, providing the potential for real-time reconstruction of occluded objects without resorting to an explicit model of the physical environment and the underlying light transport process.

In our experiments, handwritten digits from the MNIST database were displayed on an occluded laptop screen, which served as a self-luminous object. The scattered light off a nearby surface was captured by a camera on the other side of the occluder (Fig.1a). We considered four scattering surfaces ($S_1$: painted wall, $S_2$: #60 sandpaper, $S_3$: matte plastic sheet, and $S_4$: bubble wrap) with varying reflection properties. $S_1$ and $S_2$ exhibit a cosine angular dependence, characteristic of Lambertian surfaces, leading to isotropic diffuse reflection. On the other hand, $S_3$ and $S_4$ both have a pronounced specular component on top of a broad Lambertian background. $S_4$ also shows significant dependence on illumination position due to the non-uniform surface structures (see measured bidirectional reflectance distribution functions at near normal incidence in Figure S5).

A significant challenge of NLOS imaging is that an explicit physical model relating the object to the measurement is oftentimes not readily available. Here we demonstrate a NLOS imaging method through fully connected deep neural network (DNN). The DNN architecture (see Methods) can approximate the underlying physical processes without actually resorting to explicit modeling of the light transport process or the material properties of scattering surfaces. Instead, it provides a general model with adjustable parameters to abstract characteristics unique to any NLOS imaging system.

Experimental datasets were first generated for training and testing the DNN. For each scattering surface, we gathered 60,000 training scattering patterns (corresponding to the MNIST training dataset) and 10,000 testing scattering patterns (corresponding to the MNIST testing dataset). Representative results are given in Fig. 1(b)-(f), showing typical handwritten digits 0-9 from the MNIST testing data (ground truth), the corresponding measured scattering patterns from each of the scattering surfaces, and the reconstructed images. The reconstruction from $S_1$ and $S_2$ is relatively blurry, possessing less details of edges and fine features. Nevertheless, the DNN can still reconstruct images of untrained digits from the testing dataset.

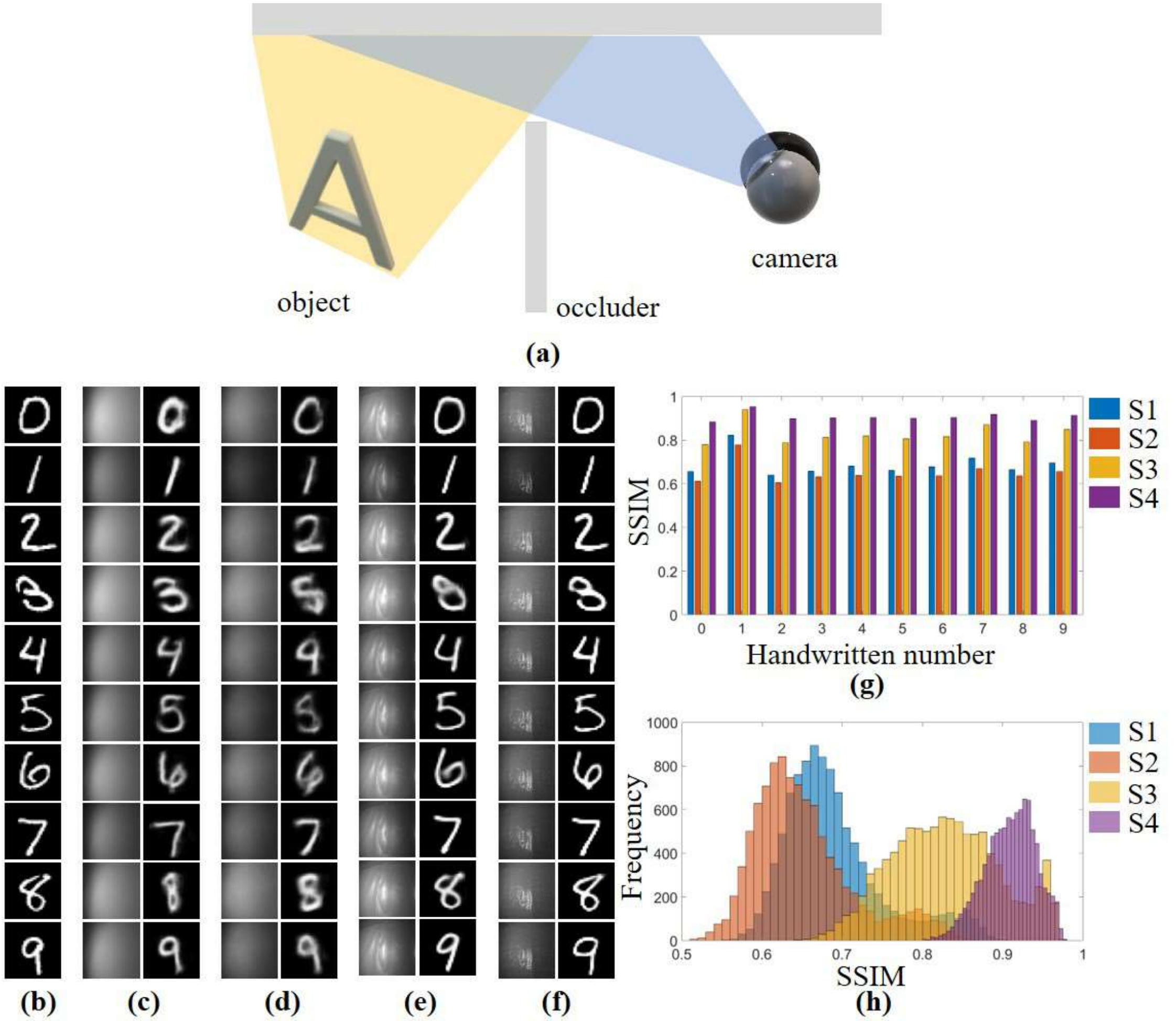

*Figure 1. (a) Schematic diagram of the NLOS experimental setup. An object is positioned on one side of an occuler. The scattered light off a scattering surface is captured by a camera on the other side of the occuler. (b)-(h) NLOS imaging results. A DNN is trained for each of the four scattering surfaces (S1, S2, S3, S4) using scattering patterns produced by the handwritten digits in the MNIST training dataset, and tested using scattering patterns produced by the MNIST testing dataset. (b): representative testing digits from the MNIST testing dataset (ground truth); (c) left: captured scattering patterns produced by the scattering surface S1, right: images reconstructed by using the trained DNN from the corresponding scattering patterns shown on the left column; (d)-(f): scattering patterns produced by S2, S3, and S4, respectively, and the corresponding reconstruction results; (g): structural similarity (SSIM) for each digit obtained from S1 to S4, and (h) is the histogram plot of SSIM for S1 to S4. These results show that NLOS imaging using S4 yields the best reconstruction results and the highest SSIM.*

In comparison, $S_4$ produces the best reconstruction results that match the ground truth, despite the fact that the captured scattering patterns do not resemble any of the handwritten digits at all. To quantitatively evaluate the reconstruction quality, we computed the structural similarity index (SSIM)[23] corresponding to each of the 10,000 testing data and for each of the scattering surfaces. The bar chart for each digit and the histograms are presented in Fig. 1 (g)-(h). For Lambertian surfaces $S_1$ and $S_2$, the SSIMs of reconstructed images are about 0.63. The SSIMs obtained from $S_3$ are around 0.83, and that from $S_4$ center around 0.93. These results indicate that a scattering surface with larger specular contribution tends to yield higher reconstruction fidelity. Higher reconstruction quality implies that the trained DNN model better approximates the underlying physics, and may exhibit generalized applicability to imaging objects increasingly distinct from these used to train the very model.

We experimentally examined the generalization of the DNN model trained for S4, which yields the highest SSIM. We began by studying the tolerance to the shift and rotation of a test object without additional training. A digit from the MNIST testing database was shifted over a distance of ±3.25 cm in lateral directions and rotated from 0° to 330° at a step of 30°. The captured scattering patterns were then analyzed by the DNN trained using the standard MNIST training dataset as described earlier. The results in Fig. 2 show that the reconstructed images faithfully trace the shift and the rotation of the input digit, indicating our DNN can generalize to reconstruct patterns independent of the lateral displacement and orientation of an input object. The complete shifting and rotation testing results are shown in Figure S7.

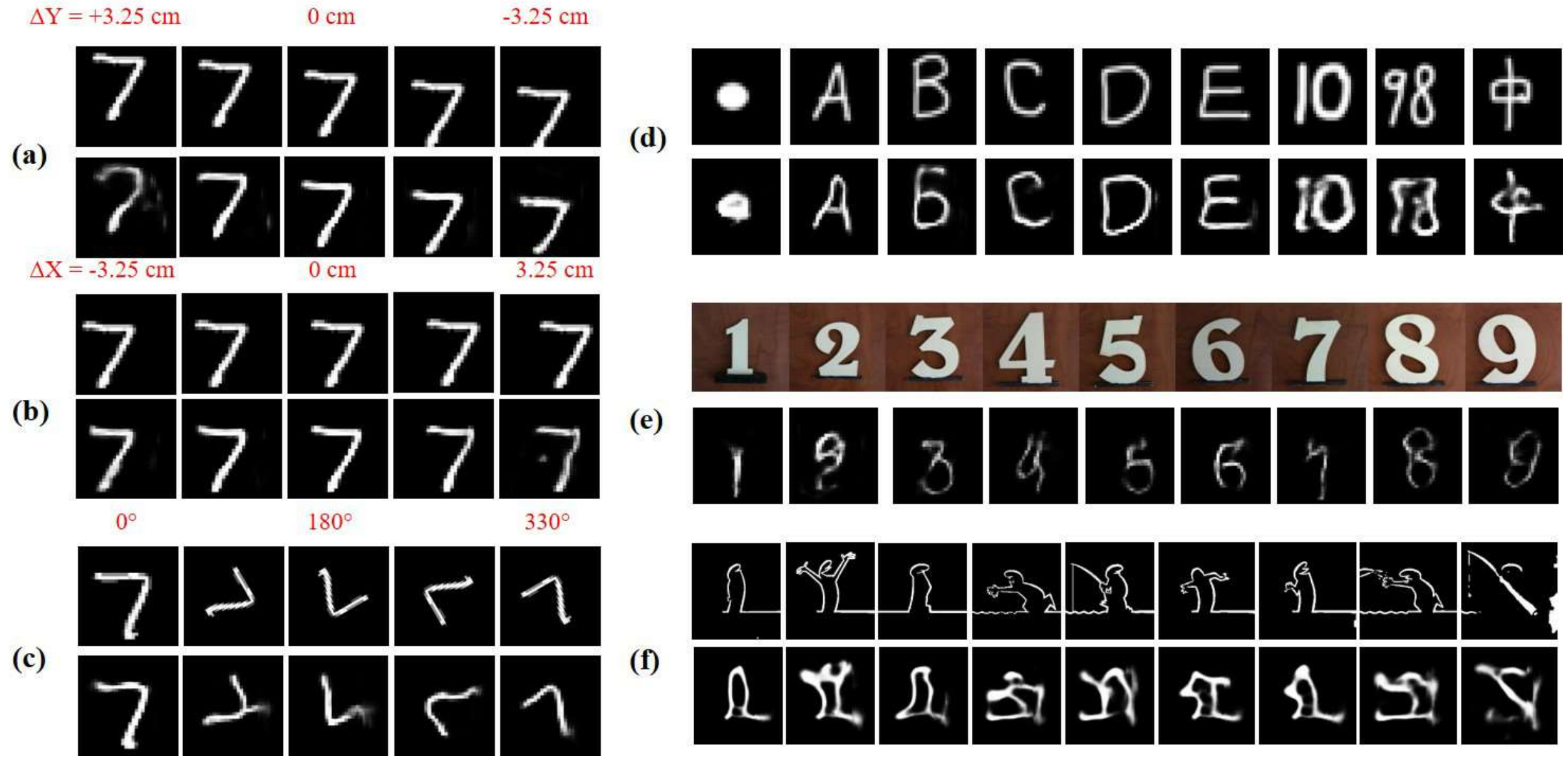

*Figure 2. (a)-(c) Shifting and rotation test of NLOS imaging system with a representative number 7 from MNIST testing dataset. (a) upper row: ground truth of the digit shifting in the vertical direction. Lower row: reconstruction result using the trained DNN from the captured scattering patterns. (b), (c) are the test for shifting in the horizontal direction, and rotation with respect to the center of the digit from 0° to 330°. Upper row indicates ground truth and lower row indicates reconstruction results. (d)-(f) Generalization of DNN trained using only the standard MNIST dataset for NLOS imaging of new patterns, real physical objects, and cartoon video - (d) Upper row: some additional testing objects including a dot, handwritten English letter A to E, handwritten two-digit number 10, 98, and a Chinese character, all of which are unseen during the training process; Lower row: the reconstructed images. (e) wooden number blocks (upper row) used as input objects, and the corresponding reconstruction results (lower row); (f) video-rate, real-time NLOS imaging. Upper row: several frames from a Youtube cartoon video; lower row: reconstruction results.*

We next tested the generalization of our DNN model by attempting to reconstruct patterns that are different from the MNIST digits used to train and test the DNN, including double handwritten digits, dot, and letters. The DNN can reconstruct these patterns reasonably well as shown in Fig. 2(d), indicating that the DNN approximates the underlying physics of the NLOS imaging and learns the proper parameters for general image reconstruction. We also conducted an experiment to validate its ability to reconstruct real physical objects instead of displayed patterns. A wooden number block (approximately 3" x 6" in size) was illuminated with a halogen light source (HL-2000-LL, Ocean Optics). Similarly, the light reflected off the wooden block reached the scattering surface ($S_4$), and the resultant pattern was captured by the camera. The DNN trained using the standard MNIST database was then used to reconstruct the inputs from these captured patterns. The results are shown in Fig. 2(e). Notably, the DNN can still reconstruct images of real physical objects.

Once a DNN is trained, the model can be applied to real-time reconstruction with minimal computational cost. In our study using a desktop computer, about 50 images per second can be processed and saved, enabling real-time video recording and reconstruction through the NLOS imaging setup and the trained DNN. We fed the MNIST-trained DNN with a video of scattering patterns generated by displaying a YouTube animation[24] . Figure. 2(f) shows nine representative frames from the video. From left to right, reconstructed, captured, and the originally displayed images are shown, respectively). Each frame is distinctly different from the trained digits. Although the quality of the retrieved images is not as good as the digits, it demonstrates a proof-of-concept result of video-rate, real-time NLOS imaging. The results further demonstrate that the DNN model trained for an NLOS imaging system can generalize and reconstruct dramatically different patterns.

It is known that the geometric image information of an object is encoded in its light field,[25,26] which describes both the light intensity distribution and the directions of rays. In an NLOS imaging system, the light field is intercepted by a scattering surface, leading to distortion of the original light field (e.g., modulated light intensity and scrambled ray directions). Scattering off a Lambertian surface results in a complete loss of the original ray directions, while reflection from a surface with appreciable specular contribution can retain a significant portion of the light field information. We hypothesize that the ability to reconstruct an object with high fidelity in NLOS imaging depends on the amount of specular contribution in the surface scattering involved, or the degree of preservation of the object's original light field, as indicated by our experimental results.

To shed light on the specular contribution enhanced NLOS reconstruction, numerical studies are performed based upon an approximate forward model. We denote the local coordinates at the object and the imaging plane as $[x_o, y_o]$ and $[x_i, y_i]$, respectively, and assume that the object has an intensity distribution of $f(x_o, y_o)$. The radiant intensity incident upon the scattering surface is given by the integral of the contribution from every constituent point source of the object. The scattering surface is modeled by a bidirectional reflectance distribution function, which contains both Lambertian and specular contributions based on the Phong model.[27] The weight of each contribution is represented by $\sigma_l$ (Lambertian) and $\sigma_s$ (specular), respectively, and the angular width of the specular component is controlled by a parameter $\gamma$, also called the shininess factor. A higher $\sigma_s/\sigma_l$ ratio means that the surface tends to be more specular. $\gamma$ is larger for smoother surfaces, and approaches infinite when describing ideal mirrors. With considerations of the effective detecting area and the effective solid angle subtended by the imaging lens of the system, it can be shown that the detected signal in the imaging plane is given by

$$I(x_i, y_i) \approx A_p \pi {a_{lens}}^2 \iint dx_o dy_o f(x_o, y_o) \frac{(\vec{r_o} - \vec{r_s}) \cdot \hat{z}}{\left|\vec{r_\imath C}\right|^2 |\vec{r_o} - \vec{r_s}|^3} \left( \frac{\overrightarrow{OC} \cdot \overrightarrow{r_s C}}{|\overrightarrow{OC}||\overrightarrow{r_s C}|} \right)^2 \cdot \left( \sigma_l(x_s, y_s) + \sigma_s(x_s, y_s) \left( \frac{\vec{r_s} - \vec{r_o} + (2\vec{r_o} \cdot \hat{z})\hat{z}}{|\vec{r_o} - \vec{r_s}|} \cdot \frac{\vec{r_\imath} - \vec{r_s}}{|\vec{r_\imath} - \vec{r_s}|} \right)^\gamma \Bigg/ \left( \frac{\vec{r_\imath} - \vec{r_s}}{|\vec{r_\imath} - \vec{r_s}|} \cdot \hat{z} \right) \right)$$

eq. 1

The detailed derivation of eq. 1 and definitions of relevant parameters can be found in the Supplementary information.

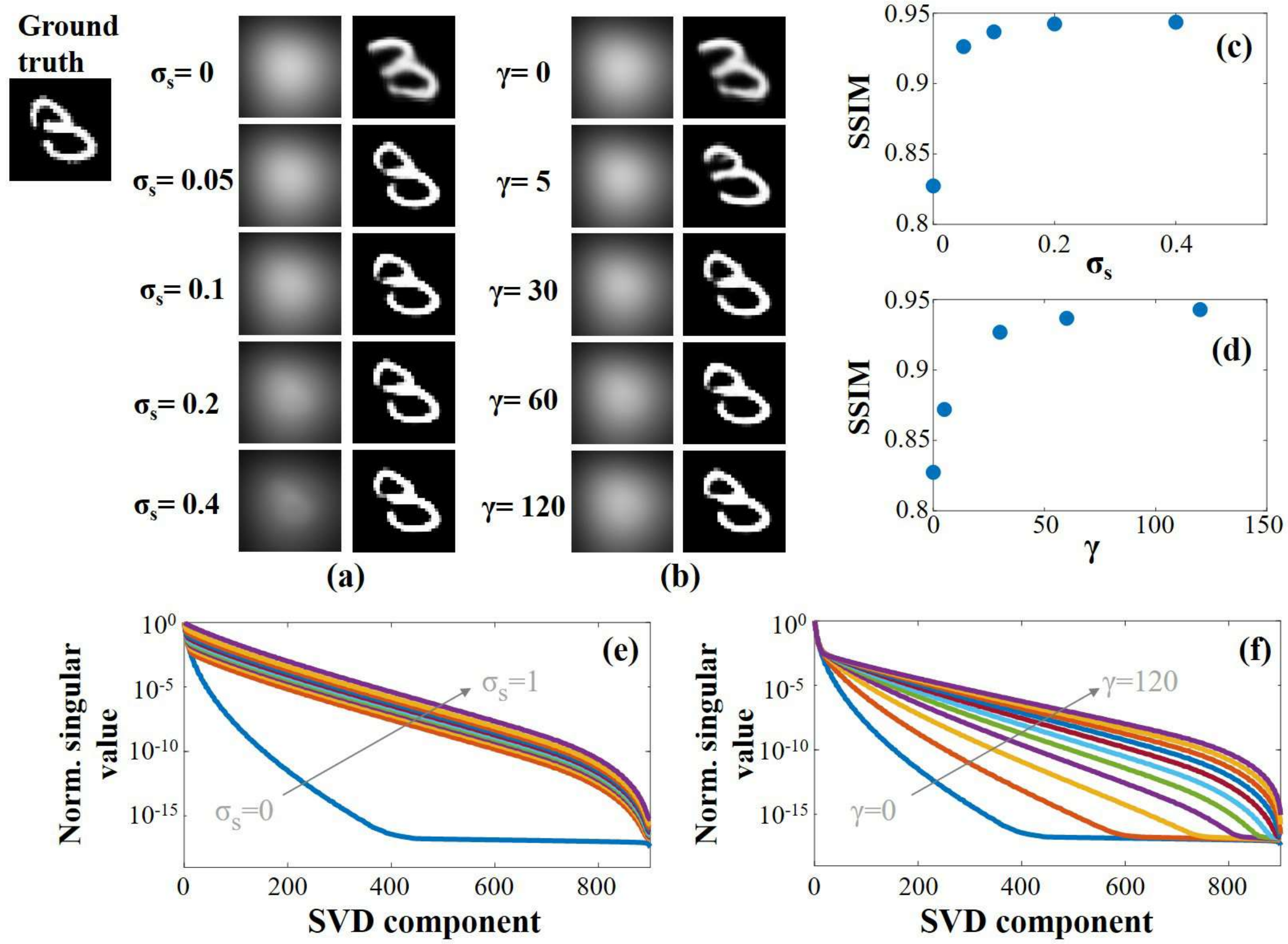

*Figure 3. Simulation results based on the proposed forward model. (a) Dependence on σs. First column: simulated scattering patterns of a handwritten digit, produced from scattering surfaces with a fixed γ value of 60 and increasing specular contribution ($\sigma_s$ increases from 0 to 0.4 while $\sigma_l$ decreases accordingly from 1 to 0.6); Second column: reconstruction results based on the simulated scattering patterns; As the specular contribution increases the reconstruction quality visibly improves. (b) Dependence on γ. First column: simulated scattering patterns of a handwritten digit, produced from scattering surfaces with a fixed $\sigma_s$ value of 0.1 and increasing shininess (γ increases from 0 to 120); Second column: reconstruction results based on the simulated scattering patterns; As the shininess increases the reconstruction quality also improves. (c-d): calculated SSIM as a function of σs and γ, respectively. (e): Normalized singular value spectrum with varying $\sigma_s$'s (γ set to 60). Larger specular contribution (i.e. larger $\sigma_s$) leads to more SVD components that are significant. (f) Normalized singular value spectrum with varying γ's ($\sigma_s$ set to 0.1). As γ increases there are more SVD components that are significant.*

Note that the Phong model does not conserve energy, and rigorously speaking, cannot be used to describe a real physical system.[28,29] Yet, due to its simplicity it provides a computationally affordable model to gain insight into the role of specular contribution in NLOS reconstruction.

Based upon the approximate forward model, we simulated our experiment by generating the training, validation, and testing scattering pattern data numerically. 60,000 MNIST digits were used as inputs to calculate the scattering patterns produced by scattering surfaces with varying $\sigma_s$ and γ in order to model the effects of surfaces with different specular properties. We considered two different ways of varying the specular contribution. In the left column of the Fig. 3(a), the γ parameter is fixed at 60 while the weighting factor of the specular contribution ($\sigma_s$) increases. A zero $\sigma_s$ represents a completely Lambertian surface, producing a diffuse reflectance pattern. When $\sigma_s$ reaches 0.4, a visible shiny bright spot, or a specular highlight, develops. In the left column of the Fig. 3(b), on the other hand, the weighting factor $\sigma_s$ is fixed at 0.1 as γ increases from 0 to 120, gradually changing from more diffusive to narrower angular distribution. Note in the simulation we assume that the entire surface is uniform and has identical $\sigma_s$ and γ values, whereas in our experiments the scattering surfaces are not spatially uniform. Nevertheless, the calculated scattering patterns share some common characteristics as the experimentally captured ones (Fig. 1). The reconstruction results based on the simulated patterns are shown in the right column of Fig. 3(a-b). As can be observed, even for the purely Lambertian surface the DNN can still reconstruct the digits with a good SSIM (~0.83). However, the reconstruction quality increases dramatically as a small amount of specular contribution is included (compare the cases of $\sigma_s$=0 and 0.05). The reconstruction also improves as $\gamma$ increases from 0 to 120, although not as visibly striking as the influence of $\sigma_s$. Our simulation indicates that for $\sigma_s > 0.05$ and $\gamma > 30$, a SSIM greater than 0.9 can be achieved and the edges of the reconstructed images are sharper, showing high reconstruction fidelity. The simulation results verify our hypothesis that the presence of specular contribution significantly improves the performance of the reconstruction, even though reasonable reconstruction for purely Lambertian surfaces is achievable.

To gain additional insight, we discretize eq. 1 and rewrite it in a matrix form $\mathrm{I_i} = \sum_{(\mathrm{x_o},\mathrm{y_o})} \mathrm{M_{io}} \mathrm{f_o}$, where $\mathrm{I_i}$ is a vector corresponding to the captured scattering pattern $\mathrm{I}(x_i, y_i)$, $\mathrm{f_i}$ is a vector corresponding to the input $\mathrm{f}(x_o, y_o)$, and $\mathrm{M_{io}}$ is a matrix that describes the integral kernel function. Intuitively, the reconstruction quality should positively correlate with the pseudo-invertibility of the M matrix, which can be visualized by performing the singular value decomposition (SVD)[30], as shown in Fig. 3(e-f). As the weight and the shininess factor increase, i.e., the specular contribution increases, the singular value spectrum broadens. The 50-dB bandwidth is increased by more than 4X with a 10% specular contribution ($\sigma_s$=0.1 and $\gamma$=60), which indicates that at the presence of even a small amount of specular contribution the NLOS measurement can retain significantly more singular value components, leading to higher reconstruction fidelity (higher SSIM).

Finally, to compare with the simulation we experimentally measured the bidirectional reflectance distribution function for each scattering surface and fitted it using the Phong model, as described in the Methods. The fitted curves are shown in dashed lines in Figure S5, and the obtained $\sigma_s$ and $\gamma$ values are tabulated in Table S1. As can be observed, $S_1$ and $S_2$ are very close to Lambertian surfaces because of the small $\gamma$ values, while $S_3$ and $S_4$ have both Lambertian and specular components and $S_4$, in particular, has stronger specular contribution compared to $S_3$ in some positions. The mirror, although in theory its $\gamma$ should be infinite, had a fitted $\gamma$ value of 147 due to the limited angular resolution of the measurement system. As discussed previously, $S_3$ and $S_4$ yield higher SSIM's than $S_1$ and $S_2$. Particularly, $S_4$ exhibits a SSIM above 0.9. This result is consistent with the simulation.

**Conclusion**

In this work a deep learning based NLOS imaging system is developed, which can "see around the corner" using a low-cost camera and incoherent light illumination. Although only trained with MNIST digit database, the DNN model can reconstruct untrained digits, other types of patterns including cartoon animation, and real objects. Once the DNN is trained, real-time reconstruction (50 frames/s) can be achieved. This study also sheds light into the critical role of specular contribution in enhancing reconstruction fidelity. Many of the difficulties of NLOS imaging lie in the fact that the light field of a hidden object is significantly distorted by surrounding scatterers upon re-directing information-carrying light rays into the detector, and thus only partial information of the hidden object is accessible. Qualitatively speaking, specular contribution can help retain the light ray propagation direction and reduce the loss of the light field information. Quantitatively, the presence of specular contribution can broaden the singular value spectral bandwidth of the relevant measurement matrix, i.e., the operator that maps the unknown object function into measured data. In other words, more significant singular value components are retained in the measured data if specular contribution is present, which enhances the pseudo-invertibility as well as learning-based reconstruction fidelity. The notion of incorporating specular reflection to retain the light field information may have different manifestations in other techniques. For example, photons scattered in the specular direction can reach the detector first (Fermat's principle[31]) and hence time-gating effectively realizes the effect of specular contribution. The presented method simplifies the experimental implementation without using high-cost and bulky lasers as well as high-speed and high-sensitivity detectors. It does not rely on explicit, precise modeling of the underlying light transport process, and is thus capable of adaptively learning many other types of scattering surfaces to reconstruct hidden objects in real time.

## Methods

*NLOS Experimental setup*

The schematic of the NLOS imaging system is shown in Fig 1 and Fig S1 (supplementary information). A laptop (Hewlett-Packard Omen, 15-inch display) acts as a self-luminous object, and displays handwritten digits from the MNIST database (60,000 training and 10,000 testing datasets, 8-bit grayscale figures, 28x28 image resolution) with a size of ~18 cm x 18 cm toward a scattering surface. The scattering patterns are then recorded by a webcam (BCC950, Logitech) located at the other side of the occluder. The object distance ($L_o$) and imaging distance ($L_i$), measured from the scattering surface, are both 20 cm, with 15° incident ($\alpha$) and detecting ($\beta$) angles. The setup is placed in a dark room to prevent any external noises that could potentially disturb the experimental measurement.

*DNN architecture and training process*

The DNN has a nearly symmetric architecture, with contracting and expanding path for downsampling and upsampling.[32] Cross-entropy loss function is used for backpropagation. Compared to commonly used loss functions, such as mean squared error or mean absolute error, the cross-entropy loss function can promote sparsity and penalize prediction error more when the predicted values deviate significantly from the ground truth. The averaged cross-entropy loss L is given by

$$L = \frac{1}{N}\sum_{i}\sum_{p}\sum_{q} -(y_i' log(y_i) + (1 - y_i')\log(1 - y_i))$$

where $y_i'$ is the ground truth pixel value of the $i^{th}$ training sample, and $y_i$ is the corresponding predicted pixel value by the DNN. $p$ and $q$ are the spatial coordinates of the image. $N$ is the total number of training images.

We resize a captured scattering pattern into a 256x256, gray scale, 8-bit image by using the bicubic interpolation method, and normalize pixel values to the range of [0,1] for training the DNN architecture (Figure S4). The output image size is 256x256, the same as the input. The optimizer that we use is the Adam method of stochastic gradient descent. We use the scattering patterns produced by the MNIST database of handwritten digits to train and test the DNN. The MNIST database consists of 60,000 training sets, and 10,000 testing sets. Additionally, we include testing images other than handwritten digits to test the performance of the DNN architecture.

The software program is written in TensorFlow 1.5 with Python 3.6. The machine used for this work is a windows-based computer with Intel i9-9900k at 3.6 GHz and 32 GB of RAM. We use NVIDIA GeForce RTX 2080 Ti (11 GB GDDR6) as the graphic processing unit to perform DNN computations. The training time is about 20 hours.

*SSIM factor*

SSIM[23] quantifies the quality of reconstructed images with reference to the ground truth, containing a luminance term, a contrast term and a structural term. An index in the range of [0,1] is calculated, with 1 being the best, and 0 the worst. The SSIM is given by

$$SSIM(x, y) = [l(x, y)]^{\alpha} \cdot [c(x, y)]^{\beta} \cdot [s(x, y)]^{\gamma},$$

where

$$l(x,y) = \frac{2\mu_x\mu_y + C_1}{\mu_x^2 + \mu_y^2 + C_1}$$

$$c(x,y) = \frac{2\sigma_x\sigma_y + C_2}{\sigma_x^2 + \sigma_y^2 + C_2}$$

$$s(x,y) = \frac{\sigma_{xy} + C_3}{\sigma_x\sigma_y + C_3}\ .$$

For the images $x\ and\ y$, $\mu\ and\ \sigma$ are the local mean and standard deviation, $\sigma_{xy}$ is the cross-covariance, and $\alpha, \beta\ and\ \gamma$ are the different weights for luminance term, the contrast term and the structural term, respectively. Here we choose $\alpha = \beta = \gamma = 1$. $C_1, C_2$ and $C_3$ are the regularization constants to avoid instability. We choose $C_1 = (0.01L)^2, C_2 = (0.03L)^2, C_3 = \frac{C_2}{2}$ , where $L$ is the dynamic range value of tested images.[23]

*Bidirectional reflectance distribution functions (BRDF) measurement*

A collimated white light beam (HL-2000-LL, Ocean Optics) was normally incident onto the test surface. We used a multi-mode fiber (M28L01, Thorlabs) to collect the scattered light at different angles from 0° to 80° at a step of 10°. A spectrometer (Flame-S-VIS-NIR, Ocean Optics) was used to measure the spectra of the scattering light. The intensities of the scattered light were obtained by integrating each spectrum. Note that because $S_4$ has highly uneven surface structure, we selected two positions indicative of different representative characteristics. $S_5$ is a mirror, used as a control to estimate the angular measurement precision. The measured data is plotted with solid dots in Figure S5. We fitted these data to the Phong model by solving a regularized optimization problem as follows.

*Fitting the Phong model with experimental measurements*

We use the following function $f(\theta)$ to fit the measured data. The loss function has regularization terms to stabilize the parameter fitting.

$$f(\theta) = \sigma_L \cos(\theta - \theta_0) + \sigma_S \cos(\theta - \theta_0)^\gamma$$

$$loss\ function = \sum_{1}^{N} (y_i - f(\theta))^2 + \alpha_1\sigma_S + \alpha_2\gamma$$

$y_i$ are the measured data points. $\sigma_S$ and $\gamma$ are the weight of the specular contribution and the shininess factor, $\sigma_L$is the weight of the Lambertian contribution, $N$is the number of points in the BRDF measurement, $\alpha_1$ and $\alpha_2$ are the weight factors for the regularization terms. Here we set $\alpha_1 = 0.01$, $\alpha_2 = 0.0005$. The regularization terms favor smaller $\sigma_S$ and $\gamma$ parameters.

*Transfer matrix* $M_{io}$ *for Singular Value Decomposition (SVD) analysis*

The matrix $M_{io}$ is the transfer matrix that transforms an input object $f_o$ to its scattering pattern $I_i$ (i.e., $I_i = M_{io}f_o$). Both $f_o$ and $I_i$ are represented as column vectors, with a size of B×1 and A×1, respectively. $M_{io}$

is thus an A-by-B matrix (A=$128^2$ and B=$30^2$ in our case), and can be expressed as $\mathrm{M_{io}} = [M_1, M_2, .., M_b, \ldots, M_B]$. Each vector $M_b$ is the corresponding impulse response generated by a point source placed at the $b^{th}$ entry in the input. For example, $M_2$ can be obtained by inputting a point source $[0,1,0, \ldots,0]^T$ into the simulation model. The schematic diagram of this process is depicted in the Figure S6.

**Acknowledgements**

This work was funded by the Penn State MRSEC, Center for Nanoscale Science, under the award NSF DMR-1420620.

**Author contributions**

C.Z., C.-Y.W., and Z.L. conceived and developed the methods. C.Z. and C.-Y.W. built the imaging system, ran the simulation, and implemented the reconstruction. Z.L. supervised the project. All authors contributed to writing the paper.

**Competing interest declaration**

The authors declare no competing interests.

Supplementary information for

# Non-line-of-sight imaging off a Phong surface through deep learning

Chen Zhou[1], Cheng-Yu Wang[1], Zhiwen Liu[*]

[1]These authors contributed equally
Department of Electrical Engineering, The Pennsylvania State University, University Park, PA, 16802, USA. *email: zzl1@psu.edu, N-211 Millennium Science Complex, The Pennsylvania State University, University Park, PA 16802

Phong surface simulation model

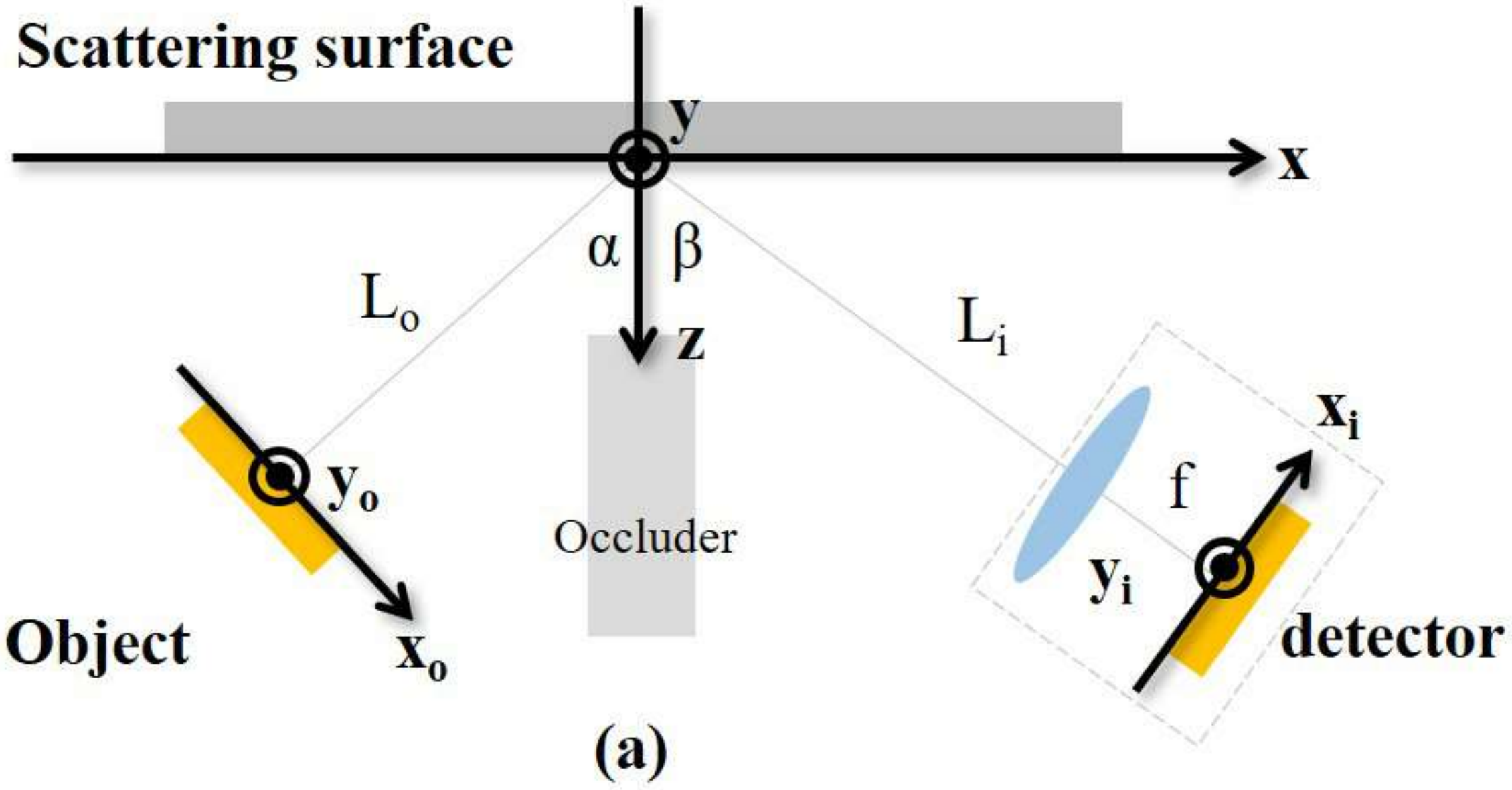

*Figure S1. Schematic of the forward model (top view)*

The relevant coordinate systems used in the simulation model are shown in figure S1. Two local coordinate systems $[x_o, y_o]$ and $[x_i, y_i]$, denoted using square brackets, are used to describe the object and detector (or imager) planes. A global coordinate system (x, y, z) is also set up with the x-y plane on the scattering surface; the origin is located where the two surface normals of the object and detector planes, each passing their respective local coordinate origins, meet with angles of α and β with respect to the z-axis, the normal direction of the scattering surface. We first express a point on the object or the detector plane in terms of the global coordinate system. Based on the configuration in figure S1, we have

$$\vec{r}_o = (x_o\cos\alpha - L_0\sin\alpha, y_o, x_o\sin\alpha + L_0\cos\alpha) \qquad \text{(S1)}$$

$$\vec{r}_i = (x_i\cos\beta + (L_i + d)\sin\beta, y_i, -x_i\sin\beta + (L_i + d)\cos\beta) \qquad \text{(S2)}$$

where $\vec{r}_o$ and $\vec{r}_i$ denote points with local coordinates $[x_o, y_o]$ and $[x_i, y_i]$ on the object and the detector planes, respectively, under the global coordinate system, $L_o$ is the distance between the global origin and the local origin on the object plane, $L_i$ is the distance between the center of the imaging lens and the global origin, and d is the distance between the lens and the detector plane. For simplicity, the optical axis of the lens is assumed to pass both the global origin and the local origin on the detector plane. We further assume that the imaging lens has a large enough depth of field so that every detector pixel centered at $\vec{r}_i$ corresponds to a specific detecting area centered at $\vec{r}_s = (x_s, y_s, 0)$ on the scattering surface, $\vec{r}_s$ can be obtained by solving the following equation that describes the chief ray passing through the center of the lens:

$$\frac{x_s - L_i\sin\beta}{x_i\cos\beta + d\sin\beta} = \frac{y_s}{y_i} = \frac{0 - L_i\cos\beta}{-x_i\sin\beta - d\cos\beta} \qquad \text{(S3)}$$

Therefore, $\vec{r}_s$ is given by

$$\vec{r}_s = \left(\frac{L_i x_i}{x_i \sin\beta - d\cos\beta}, \frac{L_i \cos\beta y_i}{x_i \sin\beta - d\cos\beta}, 0\right) \qquad \text{(S4)}$$

Now, let us consider a point source placed at $[x_o, y_o]$ on the object plane with an isotropic radiance. The radiant flux density $I(\overrightarrow{r_s})$ on the scattering surface can be expressed as (up to a constant factor) [1]

$$I(\overrightarrow{r_s}) = \frac{1}{|\overrightarrow{r_o} - \overrightarrow{r_s}|^3}(\overrightarrow{r_o} - \overrightarrow{r_s}) \cdot \hat{z} \qquad \text{(S5)}$$

This pattern is then scattered by the scattering surface and the scattered light is subsequently recorded by the detector. To simplify the model and be consistent with the experiment, we assume that the lens subtends a small solid angle from each point on the scattering surface. Let us consider a detector pixel centered at $[x_i, y_i]$ on the detector plane looking at a corresponding area *A* centered at point $(x_s, y_s, 0)$ on the scattering surface. The radiant flux which falls within the solid angle $\Omega_{lens}$, defined by the lens aperture, may be expressed in the form[1]

$$h(x_i, y_i) = \iint_{\Omega_{lens}} \iint_{A} d\Omega dx'_S dy'_S I\left(\overrightarrow{r'_s}\right) R\left(\overrightarrow{r'_s}\right) \qquad \text{(S6)}$$

where $\overrightarrow{r'_s} = (x'_S, y'_S, 0)$ and $R\left(\overrightarrow{r'_s}\right)$ is the bidirectional reflectance distribution function (BRDF) at point $\overrightarrow{r'_s}$ on the scattering surface. For simplicity, we use the Phong model[2] to describe the scattering property of the scattering surface:

$$R(\overrightarrow{r_s}) = \sigma_l(\overrightarrow{r_s})\left(\frac{\overrightarrow{r_\iota} - \overrightarrow{r_s}}{|\overrightarrow{r_\iota} - \overrightarrow{r_s}|} \cdot \hat{z}\right) + \sigma_s(\overrightarrow{r_s})\left(\frac{\overrightarrow{r_s} - \overrightarrow{r_o} + (2\overrightarrow{r_o} \cdot \hat{z})\hat{z}}{|\overrightarrow{r_o} - \overrightarrow{r_s}|} \cdot \frac{\overrightarrow{r_\iota} - \overrightarrow{r_s}}{|\overrightarrow{r_\iota} - \overrightarrow{r_s}|}\right)^{\gamma} \qquad \text{(S7)}$$

Note that $\left(\frac{\overrightarrow{r_\iota} - \overrightarrow{r_s}}{|\overrightarrow{r_\iota} - \overrightarrow{r_s}|} \cdot \hat{z}\right)$ is the cosine of the scattering angle between the surface normal ($\hat{z}$) and the scattering direction defined as from $\overrightarrow{r_s}$ to $\overrightarrow{r_\iota}$, and $\left(\frac{\overrightarrow{r_s} - \overrightarrow{r_o} + (2\overrightarrow{r_o} \cdot \hat{z})\hat{z}}{|\overrightarrow{r_o} - \overrightarrow{r_s}|} \cdot \frac{\overrightarrow{r_\iota} - \overrightarrow{r_s}}{|\overrightarrow{r_\iota} - \overrightarrow{r_s}|}\right)$ is the cosine of the angular deviation of the scattering direction from the specular reflection direction. $\sigma_l(\overrightarrow{r_s})$ and $\sigma_s(\overrightarrow{r_s})$ are the weight factors for Lambertian scattering and specular reflection, respectively. $\gamma$ is the shininess factor that specifies the "goodness" of specular reflection; for mirror-like ideal specular reflection $\gamma$ approaches to infinity.

Under the assumptions of small area A and small solid angle $\Omega_{lens}$, we can simplify eq.S6 as

$$h(x_i, y_i) = \Omega(x_s, y_s, 0) A_{deff}(x_s, y_s, 0) I(x_s, y_s, 0) R(x_s, y_s, 0) \qquad \text{(S8)}$$

where $A_{deff}(x_s, y_s, 0) = \iint_A dx'_S dy'_S$ is the effective area around the point $(x_s, y_s, 0)$ that can be detected by the detector pixel centered at $[x_i, y_i]$ on the detector plane, and $\Omega(x_s, y_s, 0) = \iint_{\Omega_{lens}} d\Omega$ represents the effective solid angle subtended by the lens. The detailed derivations of these parameters are given in the following.

**Solid angle $\Omega(x_s, y_s, 0)$**

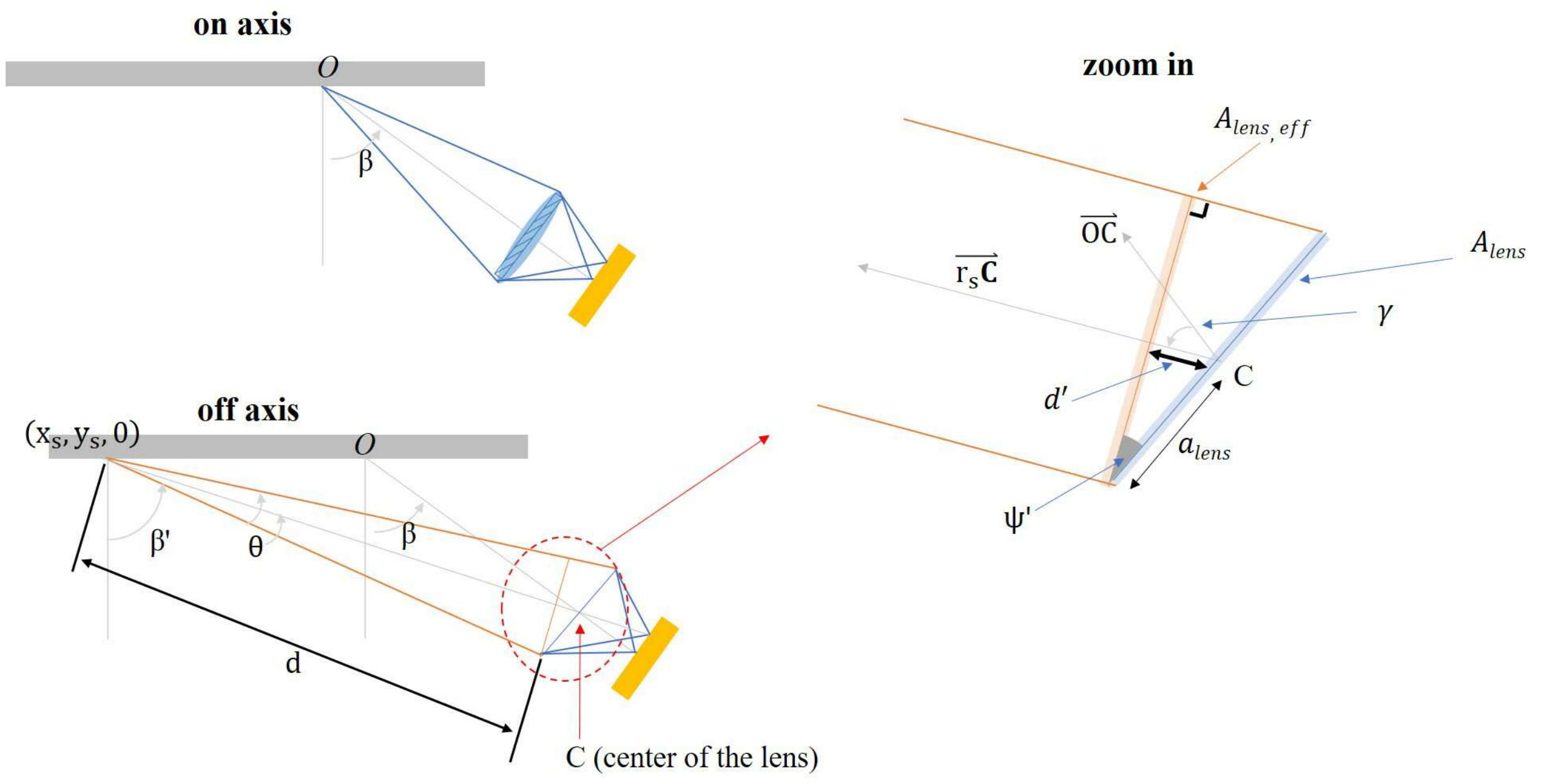

*Figure S2. Schematic for solid angle calculation*

The pixel centered at $[x_i, y_i]$ on the detector plane collects the light scattered off an area centered at the point $\overrightarrow{r_s} = (x_s, y_s, 0)$ within a solid angle Ω, which is determined by the effective lens area $A_{lens,eff}$ that is normal to the vector $\overrightarrow{r_s\mathbf{C}} \equiv \overrightarrow{OC} - \overrightarrow{r_s}$ (where $\overrightarrow{OC}$ is the position vector of the center of the lens, O is the origin of the global coordinate system, C is the center of the lens) and the distance d between the center of the effective area and the point $(x_s, y_s, 0)$, as shown in figure S2. We can express d as

$$d = \left|\,\overrightarrow{OC} - \overrightarrow{r_s}\right| - d' = \left|\overrightarrow{r_s\mathbf{C}}\right| - a_{lens}\sin(\psi') \;\approx \left|\overrightarrow{r_s\mathbf{C}}\right| - a_{lens}\sin(\psi) \;\approx \left|\overrightarrow{r_s\mathbf{C}}\right| \qquad \text{(S9)}$$

Where d' is the distance between C and the center of the effective area, $a_{lens}$ is the radius of the lens aperture, $\psi \equiv \cos^{-1}\left(\frac{\overrightarrow{OC}\cdot\overrightarrow{r_sC}}{|\overrightarrow{OC}||\overrightarrow{r_sC}|}\right)$ is the angle between vector $\overrightarrow{OC}$ and $\overrightarrow{r_s\mathbf{C}}$ (figure S2) which can approximate the angle ψ'. To be consistent with the experimental setup, here we assume that the lens area ($A_{lens}$) and its numerical aperture ($a_{lens}/L_i = 0.005$ in the experiment) are small, and thus we can approximate the marginal rays to be parallel to each other (see figure S2(c)). Based on this approximation, we can express the effective lens area $A_{lens,eff}$ as the projection of the lens area $A_{lens}$:

$$A_{lens,eff} = A_{lens}\cos(\psi) \qquad \text{(S10)}$$

We can therefore obtain the solid angle as follows

$$\Omega \approx \frac{A_{lens}\cos(\psi)}{\left[\left|\overrightarrow{r_sC}\right| - a_{lens}\sin(\psi)\right]^2} = \frac{\pi {a_{lens}}^2\left(\frac{\overrightarrow{OC}\cdot\overrightarrow{r_sC}}{|\overrightarrow{OC}||\overrightarrow{r_sC}|}\right)}{\left[\left|\overrightarrow{r_sC}\right| - a_{lens}\sin\left(\cos^{-1}\left(\frac{\overrightarrow{OC}\cdot\overrightarrow{r_sC}}{|\overrightarrow{OC}||\overrightarrow{r_sC}|}\right)\right)\right]^2} \approx \pi {a_{lens}}^2\left(\frac{\overrightarrow{OC}\cdot\overrightarrow{r_sC}}{|\overrightarrow{OC}||\overrightarrow{r_sC}|^3}\right) \qquad \text{(S11)}$$

**Effective detection area $A_{deff}(x_s, y_s, 0)$**

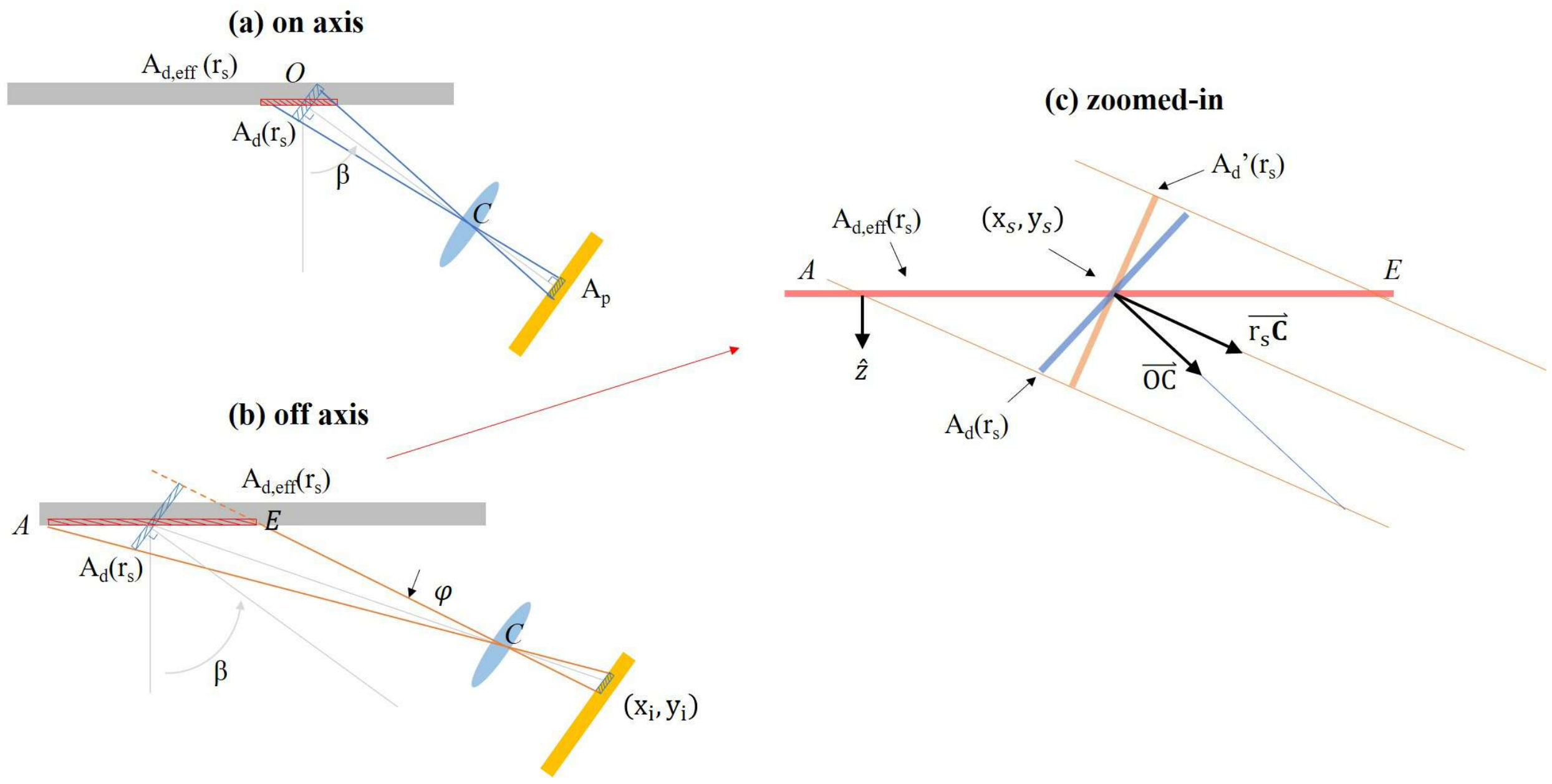

*Figure S3. Schematic for acceptance area calculation*

In our calculation, the effective detection area ($A_{deff}$ in eq.S8) takes into account the tilted detection configuration, as shown in the figure S3(a). Let us consider a pixel with an area of $A_p$ on the detector, and its corresponding image with an area of $A_d$, which has a surface normal parallel to the optical axis of the lens, i.e. the vector $\overrightarrow{OC}$ (figure 3(a)). Note that $A_d = {A_p}/{M^2}$, and $M = {|\overrightarrow{r_i \mathbf{C}}|}/{|\overrightarrow{r_s \mathbf{C}}|}$ is the magnification of the lens. Here we assume the imaging system has a sufficient depth of field and the detector pixel can collect light coming from an effective area $A_{d,eff}$ on the scattering surface. In the on-axis case where the detector pixel is centered at the local origin of the detector plane (figure S3(a)), $A_{d,eff}$ is given by $A_d/\cos\beta$ (β is defined in the first section; also see figure S2). However, in the off-axis case, where the detector pixel is not located at the center (figure S3(b)), the expression of $A_{d,eff}$ needs to be modified accordingly. Since the distance between the lens and the detector is much larger than the pixel size of the detector, we assume that the collection angle φ of each detector pixel is small (see figure S3(b)). Thus, we can assume that the chief rays ($\overline{EC}$, $\overline{AC}$) on the collection cone in the off-axis case are parallel to each other. Figure S3(c) shows an enlarged schematic under the small φ approximation. We introduce a variable $A_d'$ for representing the projection of the $A_d$ on the plane that has the surface normal of $\overrightarrow{r_s \mathbf{C}}$, so that $A_{d,eff}$ can be obtained by projecting this $A_d'$onto the scattering surface. Since the surface normal of the $A_{d,eff}$is $\hat{z}$, and the surface normal of $A_d$ is vector $\overrightarrow{OC}$ (defined by the on-axis case, figure S3(a)), we can express the $A_{d,eff}$ by:

$$A_{d,eff} = \frac{A_d{}'}{\cos\left(\measuredangle\left(\overrightarrow{r_s\mathbf{C}}, \hat{z}\right)\right)} = \frac{A_d \cos\left(\measuredangle\left(\overrightarrow{r_s\mathbf{C}}, \overrightarrow{OC}\right)\right)}{\cos\left(\measuredangle\left(\overrightarrow{r_s\mathbf{C}}, \hat{z}\right)\right)} = A_d \frac{\left(\frac{\overrightarrow{OC}\cdot\overrightarrow{r_s\mathbf{C}}}{\left|\overrightarrow{OC}\right|\left|\overrightarrow{r_s\mathbf{C}}\right|}\right)}{\left(\frac{\overrightarrow{r_s\mathbf{C}}}{\left|\overrightarrow{r_s\mathbf{C}}\right|}\cdot\hat{z}\right)}$$

$$= A_p \frac{\left|\overrightarrow{r_s\mathbf{C}}\right|^2}{\left|\overrightarrow{r_\imath\mathbf{C}}\right|^2} \frac{\left(\frac{\overrightarrow{OC}\cdot\overrightarrow{r_s\mathbf{C}}}{\left|\overrightarrow{OC}\right|\left|\overrightarrow{r_s\mathbf{C}}\right|}\right)}{\left(\frac{\overrightarrow{r_s C}}{\left|\overrightarrow{r_s\mathbf{C}}\right|}\cdot\hat{z}\right)} \tag{S12}$$

where $\measuredangle(a, \mathrm{b})$ denotes the angle between two vectors a and b.

Now, we can substitute eq.S5, eq.S7, eq.S11, and eq.S12 into eq.S8 to obtain the point spread function

$$h(x_i, y_i) \approx \frac{1}{|\overrightarrow{r_o} - \overrightarrow{r_s}|^3}\left[(\overrightarrow{r_o} - \overrightarrow{r_s})\cdot\hat{z}\right] \frac{A_p\left|\overrightarrow{r_s\mathbf{C}}\right|^2 \pi a_{lens}{}^2 \left(\frac{\overrightarrow{OC}\cdot\overrightarrow{r_s C}}{\left|\overrightarrow{OC}\right|\left|\overrightarrow{r_s C}\right|}\right)^2}{\left|\overrightarrow{r_\iota\mathbf{C}}\right|^2 \left(\frac{\overrightarrow{r_s\mathbf{C}}}{\left|\overrightarrow{r_s\mathbf{C}}\right|}\cdot\hat{z}\right)\left[\left|\overrightarrow{r_s C}\right| - a_{lens}\sin\left(\cos^{-1}\left(\frac{\overrightarrow{OC}\cdot\overrightarrow{r_s C}}{\left|\overrightarrow{OC}\right|\left|\overrightarrow{r_s C}\right|}\right)\right)\right]^2}$$
$$\cdot\left(\sigma_l(x_s, y_s)\left(\frac{\overrightarrow{r_\imath} - \overrightarrow{r_s}}{|\overrightarrow{r_\imath} - \overrightarrow{r_s}|}\cdot\hat{z}\right) + \sigma_s(x_s, y_s)\left(\frac{\overrightarrow{r_s} - \overrightarrow{r_o} + (2\overrightarrow{r_o}\cdot\hat{z})\hat{z}}{|\overrightarrow{r_o} - \overrightarrow{r_s}|}\cdot\frac{\overrightarrow{r_\imath} - \overrightarrow{r_s}}{|\overrightarrow{r_\imath} - \overrightarrow{r_s}|}\right)^\gamma\right) \tag{S13}$$

Note that the $\left(\frac{\overrightarrow{r_s\mathbf{C}}}{\left|\overrightarrow{r_s\mathbf{C}}\right|}\cdot\hat{z}\right)$ and $\left(\frac{\overrightarrow{r_\imath}-\overrightarrow{r_s}}{|\overrightarrow{r_\imath}-\overrightarrow{r_s}|}\cdot\hat{z}\right)$ in the eq.S13 are the same, because $(x_s, y_s, 0)$, C, and $[x_i, y_i]$ are on the same line, i.e., $\overrightarrow{r_s\mathbf{C}}$ is parallel to $(\overrightarrow{r_\imath} - \overrightarrow{r_s})$. Thus we can rewrite the eq.S13 to

$$h(x_i, y_i) \approx \frac{1}{|\overrightarrow{r_o} - \overrightarrow{r_s}|^3}\left[(\overrightarrow{r_o} - \overrightarrow{r_s})\cdot\hat{z}\right] \frac{A_p\left|\overrightarrow{r_s\mathbf{C}}\right|^2 \pi a_{lens}{}^2 \left(\frac{\overrightarrow{OC}\cdot\overrightarrow{r_s C}}{\left|\overrightarrow{OC}\right|\left|\overrightarrow{r_s C}\right|}\right)^2}{\left|\overrightarrow{r_\iota\mathbf{C}}\right|^2 \left[\left|\overrightarrow{r_s C}\right| - a_{lens}\sin\left(\cos^{-1}\left(\frac{\overrightarrow{OC}\cdot\overrightarrow{r_s C}}{\left|\overrightarrow{OC}\right|\left|\overrightarrow{r_s C}\right|}\right)\right)\right]^2}$$
$$\cdot\left(\sigma_l(x_s, y_s) + \sigma_s(x_s, y_s)\left(\frac{\overrightarrow{r_s} - \overrightarrow{r_o} + (2\overrightarrow{r_o}\cdot\hat{z})\hat{z}}{|\overrightarrow{r_o} - \overrightarrow{r_s}|}\cdot\frac{\overrightarrow{r_\imath} - \overrightarrow{r_s}}{|\overrightarrow{r_\imath} - \overrightarrow{r_s}|}\right)^\gamma \Big/ \left(\frac{\overrightarrow{r_\imath} - \overrightarrow{r_s}}{|\overrightarrow{r_\imath} - \overrightarrow{r_s}|}\cdot\hat{z}\right)\right)$$

$$\approx A_p \pi a_{lens}{}^2 \frac{(\overrightarrow{r_o} - \overrightarrow{r_s})\cdot\hat{z}}{\left|\overrightarrow{r_\iota\mathbf{C}}\right|^2 |\overrightarrow{r_o} - \overrightarrow{r_s}|^3}\left(\frac{\overrightarrow{OC}\cdot\overrightarrow{r_s C}}{\left|\overrightarrow{OC}\right|\left|\overrightarrow{r_s C}\right|}\right)^2$$
$$\cdot\left(\sigma_l(x_s, y_s) + \sigma_s(x_s, y_s)\left(\frac{\overrightarrow{r_s} - \overrightarrow{r_o} + (2\overrightarrow{r_o}\cdot\hat{z})\hat{z}}{|\overrightarrow{r_o} - \overrightarrow{r_s}|}\cdot\frac{\overrightarrow{r_\imath} - \overrightarrow{r_s}}{|\overrightarrow{r_\imath} - \overrightarrow{r_s}|}\right)^\gamma \Big/ \left(\frac{\overrightarrow{r_\imath} - \overrightarrow{r_s}}{|\overrightarrow{r_\imath} - \overrightarrow{r_s}|}\cdot\hat{z}\right)\right)$$

(S14)

Therefore, for an object with an intensity distribution $f(x_o, y_o)$, produced by incoherent light illumination or emission, the detected pattern on the detector plane is:

$$
\begin{aligned}
I_d(x_i, y_i) \approx & \iint dx_o dy_o \frac{f(x_o, y_o)}{|\overrightarrow{r_o} - \overrightarrow{r_s}|^3} [(\overrightarrow{r_o} - \overrightarrow{r_s}) \cdot \hat{z}] \frac{A_p \left|\overrightarrow{r_s \mathbf{C}}\right|^2 \pi {a_{lens}}^2 \left( \frac{\overrightarrow{OC} \cdot \overrightarrow{r_s C}}{|\overrightarrow{OC}||\overrightarrow{r_s C}|} \right)^2}{\left|\overrightarrow{r_1 \mathbf{C}}\right|^2 \left[ |\overrightarrow{r_s c}| - a_{lens} \sin\left( \cos^{-1}\left( \frac{\overrightarrow{OC} \cdot \overrightarrow{r_s C}}{|\overrightarrow{OC}||\overrightarrow{r_s C}|} \right) \right) \right]^2} \\
& \cdot \left( \sigma_l(x_s, y_s) + \sigma_s(x_s, y_s) \left( \frac{\overrightarrow{r_s} - \overrightarrow{r_o} + (2\overrightarrow{r_o} \cdot \hat{z})\hat{z}}{|\overrightarrow{r_o} - \overrightarrow{r_s}|} \cdot \frac{\overrightarrow{r_1} - \overrightarrow{r_s}}{|\overrightarrow{r_1} - \overrightarrow{r_s}|} \right)^{\gamma} \Big/ \left( \frac{\overrightarrow{r_1} - \overrightarrow{r_s}}{|\overrightarrow{r_1} - \overrightarrow{r_s}|} \cdot \hat{z} \right) \right) \\
\approx & \, A_p \pi {a_{lens}}^2 \iint dx_o dy_o f(x_o, y_o) \frac{(\overrightarrow{r_o} - \overrightarrow{r_s}) \cdot \hat{z}}{\left|\overrightarrow{r_1 \mathbf{C}}\right|^2 |\overrightarrow{r_o} - \overrightarrow{r_s}|^3} \left( \frac{\overrightarrow{OC} \cdot \overrightarrow{r_s C}}{|\overrightarrow{OC}||\overrightarrow{r_s C}|} \right)^2 \\
& \cdot \left( \sigma_l(x_s, y_s) + \sigma_s(x_s, y_s) \left( \frac{\overrightarrow{r_s} - \overrightarrow{r_o} + (2\overrightarrow{r_o} \cdot \hat{z})\hat{z}}{|\overrightarrow{r_o} - \overrightarrow{r_s}|} \cdot \frac{\overrightarrow{r_1} - \overrightarrow{r_s}}{|\overrightarrow{r_1} - \overrightarrow{r_s}|} \right)^{\gamma} \Big/ \left( \frac{\overrightarrow{r_1} - \overrightarrow{r_s}}{|\overrightarrow{r_1} - \overrightarrow{r_s}|} \cdot \hat{z} \right) \right)
\end{aligned}
$$

(S15)

In deriving the formula, for simplicity, we have made a small emission angle approximation and treated the emission cosine as 1.

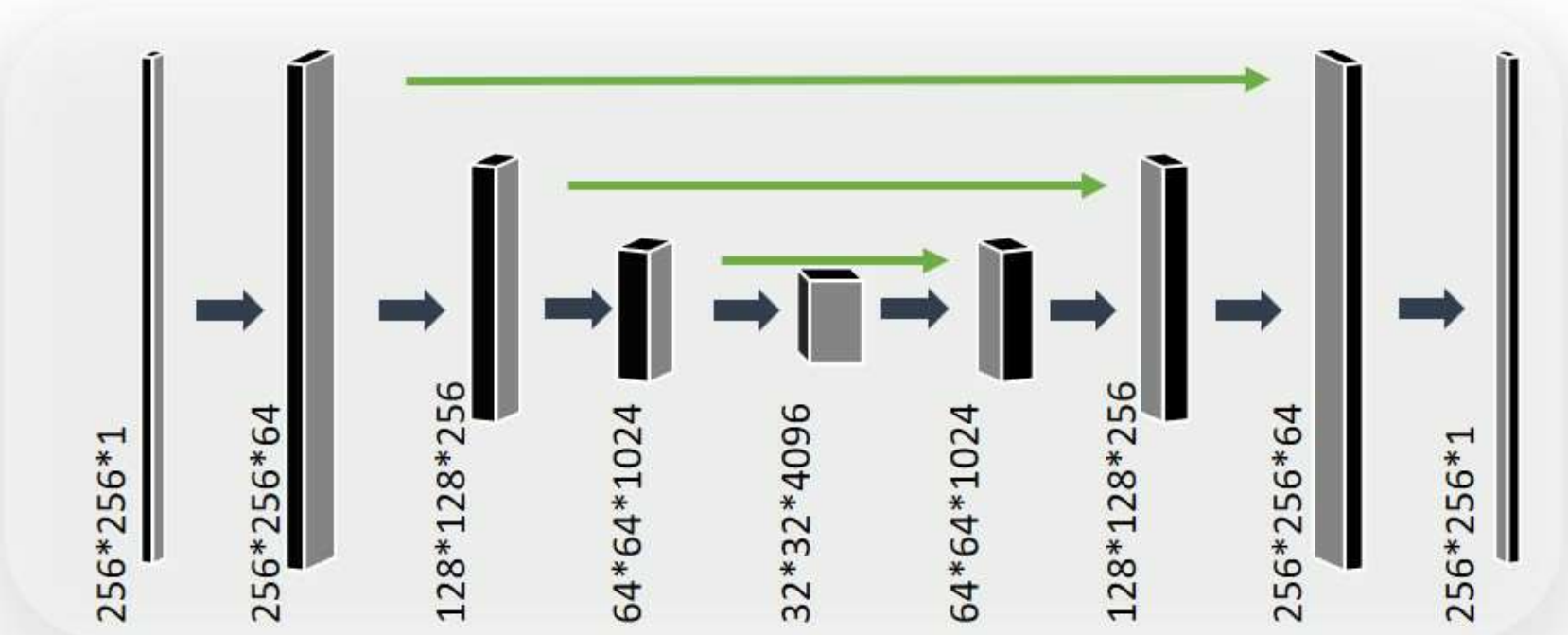

*Figure S4. DNN architecture. The inputs are scattering patterns, and the outputs are the reconstructed images. The U-Net structure consists of a downsampling path and an upsampling path for reconstruction.*

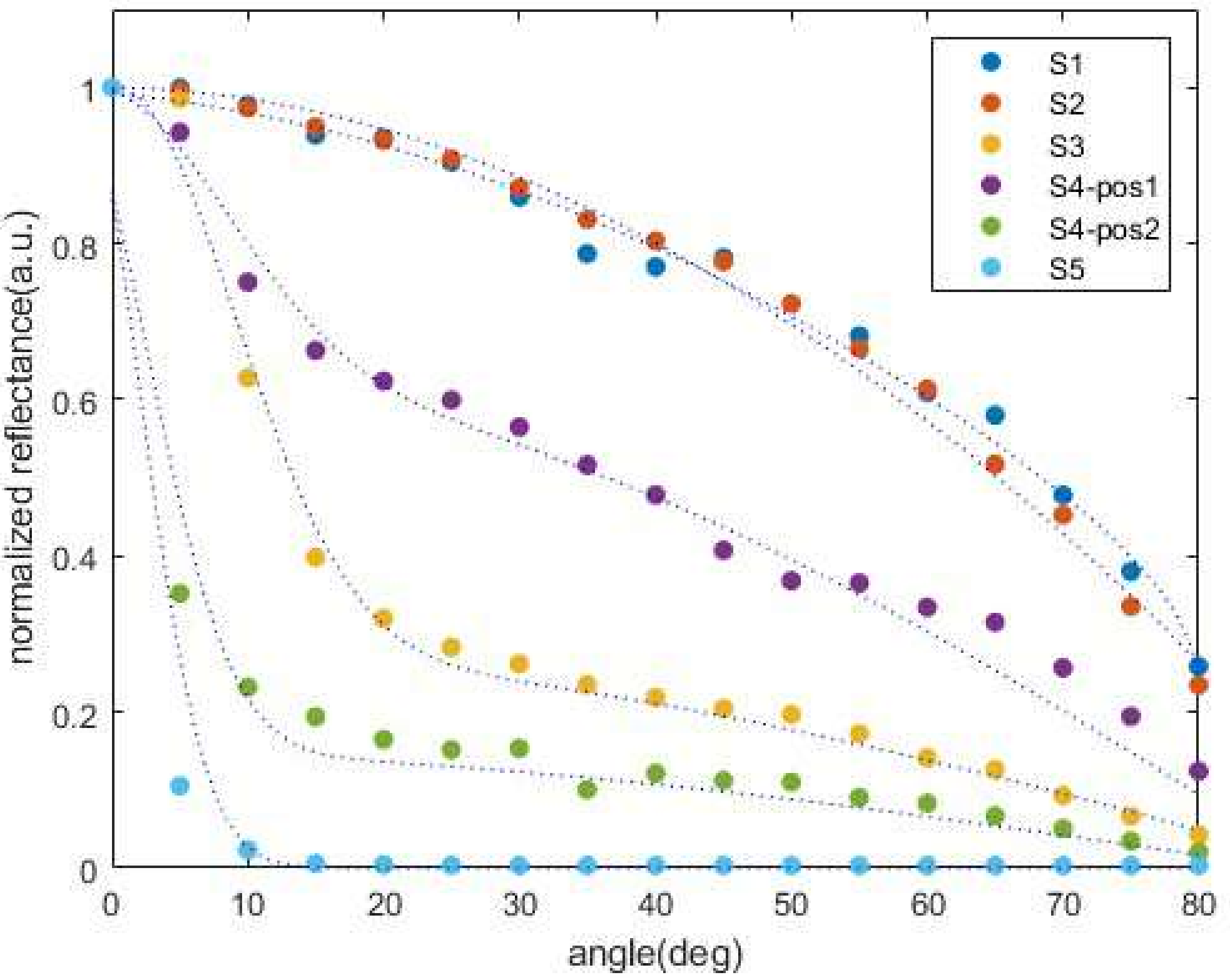

*Figure S5. Measured bidirectional reflectance distribution functions (BRDF) of several scattering surfaces used in this study and fitted curves using the Phong model. Dashed line: fitting results. Dots: experimental results.*

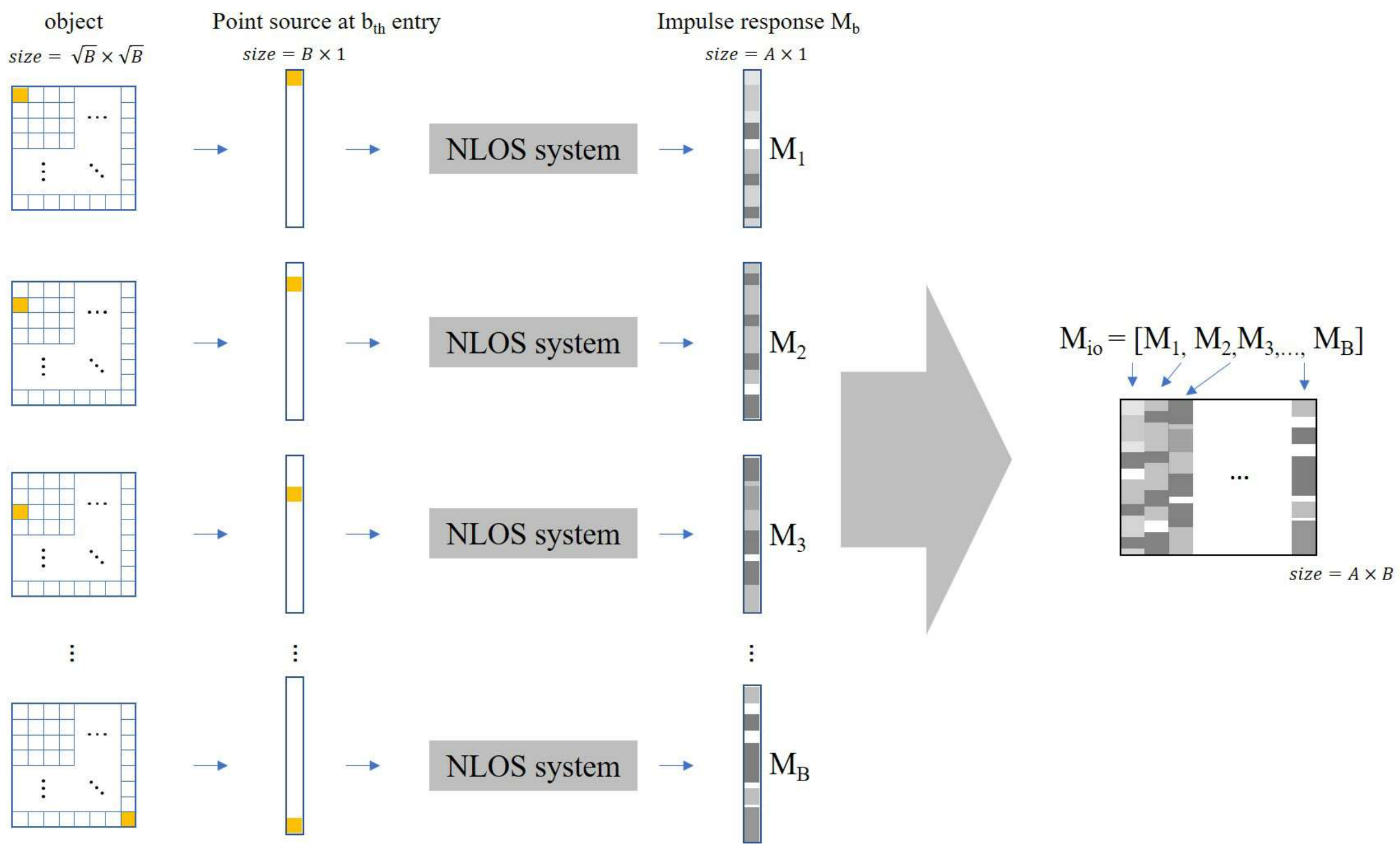

*Figure S6. Schematic of the procedure to calculate the transfer function. For each entry in the input with the size of B-by-1, we simulate the corresponding impulse response $M_b$. The final transfer matrix is the concentration of all the $M_b$'s.*

**Table S1. Fitted γ and $\sigma_s$ parameters using the Phong model (Note: $\sigma_L$ =1- $\sigma_s$)**

| | S1 | S2 | S3 | S4-pos1 | S4-pos2 | mirror |
|---|---|---|---|---|---|---|
| **γ** | 1 | 1 | 43 | 35 | 30 | 152 |
| **$\sigma_s$** | 0 | 0 | 0.57 | 0.35 | 0.80 | 1 |

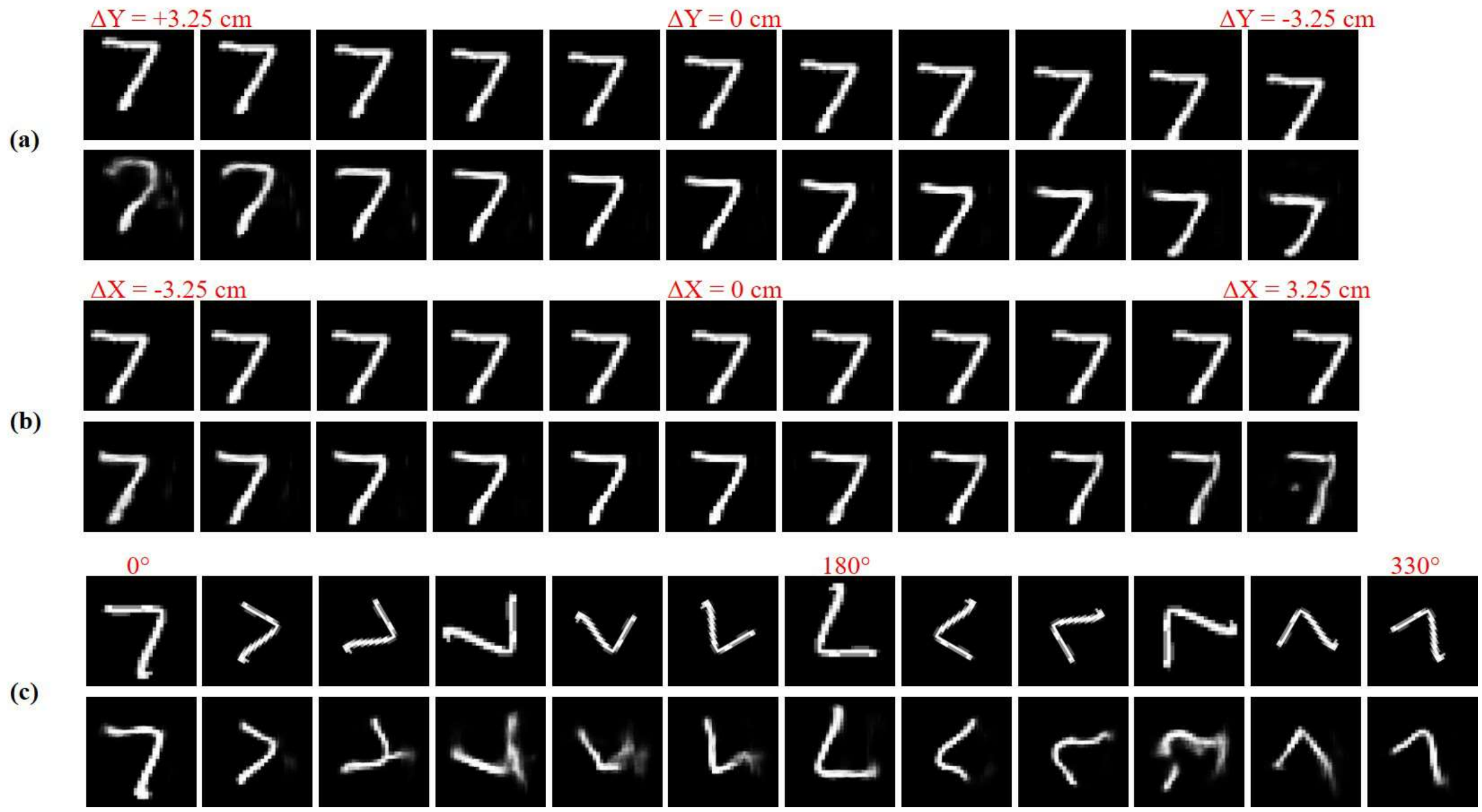

*Figure S7. Robustness of the trained DNN against shifting and rotation. (a) and (b) are the shifting test results for vertical and horizontal directions, respectively. The upper rows are the ground truth, while the lower rows are the reconstructed images using the trained DNN. The input image is shifted at a step size of 0.65 cm over a total range of ± 3.25 cm. (c) is the rotation test results. The upper row is the ground truth, and the lower row is the reconstructed images. The input image is rotated by 30° at each step from 0° to 330°.*

**<u>Parameters used in the simulation</u>**

Incident angle (α): 15 °

Distance between object and wall ($L_o$): 20 cm

Camera angle (β): 15 °

Distance between camera lens and scattering surface ($L_i$): 20 cm

Distance between camera lens and image sensor array (f): 0.8 cm

Imager size: 1 cm (=$A_p$ x 128)

Lens radius ($a_{lens}$): 0.1 cm